\documentclass[12pt]{article}
\usepackage{epsfig}
\usepackage{amsmath}

\makeatletter
\def\fmslash{\@ifnextchar[{\fmsl@sh}{\fmsl@sh[0mu]}}
\def\fmsl@sh[#1]#2{%
  \mathchoice
    {\@fmsl@sh\displaystyle{#1}{#2}}%
    {\@fmsl@sh\textstyle{#1}{#2}}%
    {\@fmsl@sh\scriptstyle{#1}{#2}}%
    {\@fmsl@sh\scriptscriptstyle{#1}{#2}}}
\def\@fmsl@sh#1#2#3{\m@th\ooalign{$\hfil#1\mkern#2/\hfil$\crcr$#1#3$}}
\makeatother
\begin{document}
%---------------- TTP Titlepage <---------------------------
\thispagestyle{empty}
\begin{titlepage}
\begin{flushright}
hep-ph/9808442 \\
TTP98--29 \\
\today
\end{flushright}

\vspace{0.3cm}
\boldmath
\begin{center}
\Large\bf One-particle Inclusive Semi-Leptonic \\ $B$ decays
\end{center}
\unboldmath
\vspace{0.8cm}

\begin{center}
{\large Christopher Balzereit} and {\large Thomas Mannel} \\
{\sl Institut f\"{u}r Theoretische Teilchenphysik,
     Universit\"at Karlsruhe,\\ D -- 76128 Karlsruhe, Germany} 
\end{center}

\vspace{\fill}

\begin{abstract}
\noindent
We propose a method for a QCD based calculation of one-particle inclusive 
decays of the form $B \to \bar D X$ or $B \to \bar D^* X$. It is based on 
the heavy mass limit and a short distance expansion
of the amplitudes, which yield a power series in the 
parameter $1/M^{2}_{X}$ for the spectra and in 
$\Lambda_{QCD}m_{b}/(m_b - m_c)^{2}$ for the 
rates. We study the leading term of this expansion for the 
case of the semi--leptonic decays $B \to \bar D X\ell^+ \nu$.     
\end{abstract}
\end{titlepage}
% ________________________________________________________________

\newpage

%%%%%%%%%%%%%%%%%%%%%%%%%%%%%%%%%%%%%%%%%%%%%%%%%%%%%%%%%%%%%%%%%%%%%%%%
\section{Introduction}
\label{sec:introduction}
Over the last ten years significant progress has been made in the 
theoretical description of heavy flavour decays \cite{reviews}. 
The application  of the $1/m_Q$ expansion ($m_Q$ being the mass 
of the heavy quark) 
allows us to perform QCD based calculations, which in some cases
yield model independent results. The additional symmetries of the 
infinite mass limit, the so called heavy quark symmetries \cite{IsgurWise}, 
reduce  the uncertainties due to unknown hadronic matrix elements 
significantly,
and corrections to this infinite mass limit have been studied extensively
using the framework of Heavy Quark Effective Theory (HQET) \cite{HQET}. 

The heavy mass expansion has been applied to various classes of decays. 
As far as exclusive decays are concerned the main progress has been 
achieved for semi--leptonic decays, while exclusive non-leptonic decays 
still have not simplified through the heavy mass limit.  

The other side are the fully inclusive decays, the rates of which can 
be obtained as a power series in $1/m_Q$ by means of an operator 
product expansion (OPE) and subsequent application of 
HQET \cite{Inclusive}. Here
semi--leptonic as well as the non--leptonic processes may be described, 
allowing us to compute lifetimes and branching ratios. The pattern 
is well reproduced by the $1/m_Q$ expansion, although some open problems 
remain \cite{NeubertHawaii}.  

Up to now no attempt has been made to apply similar methods 
to one-particle inclusive decays, such as 
$B \to \bar D X \ell^+ \nu$ or $B \to \bar D X$ and  $B \to \bar D^* X$. 
Obviously the standard method as in the inclusive case does 
not work in a naive way, since in the final state a $\bar D$ or a $\bar D^*$ 
is projected out, and the same set-up as in the inclusive case will not 
work. 

In the present paper we propose a method which allows us to compute 
one-particle inclusive rates, based on QCD. The main ingredients are 
similar as in the fully inclusive case. 
In the next section we shall describe the method 
and then discuss its application to semi--leptonic decays.

\section{Description of the Method}
We shall consider first decays of the form $B \to \bar D X$ 
(i.e. a $\bar b \to \bar c$--transition) and  
thus study the expression 
\begin{equation}
G (M^2) = \sum_X 
\left| \langle B(p_B) | H_{eff} | \bar D(p_{\bar D}) X  \rangle \right|^2
(2 \pi)^4 \delta^4 (p_B - p_{\bar D} - p_X)
\end{equation}
where $| X \rangle$ are momentum eigenstates with momentum $p_X$ and 
$H_{eff}$ the relevant part of the weak Hamiltonian. The 
function $G$ depends on the invariant mass 
$M^2 = (p_B - p_{\bar D})^2$ of the state $|X\rangle$ which ranges between 
\begin{equation}
0 \le M^2 \le (m_B - m_{D})^2\,, 
\end{equation}
where we have neglected the pion mass as well as the lepton masses.
This function $G$ is related to the decay rate under 
consideration by 
\begin{equation}
d \Gamma (B \to \bar D X) = \frac{1}{2 m_B} d\Phi_{\bar D}\,\, G(M^2) 
\end{equation}
where $d\Phi_{\bar D}$ is the phase space element of the final state $\bar D$ 
meson. 

The region close to $M^2 \approx 0$ is dominated by a few resonances 
(the $\pi$ and $\rho$ states in the non-leptonic case), and away 
from this region one can expect 
duality to hold. In particular, this should be true in the limit in which 
$m_b,\, m_c \to \infty$, since in almost all available phase space we 
have $M^2 \gg \Lambda_{QCD}$.  

In technical terms this means that we are going to set 
up a short distance expansion for the quantity $G(M^2)$. The procedure 
is similar as the one for inclusive decays, we write
\begin{equation}
G (M^2) =\sum_{X} \int d^4 x \,  
        \langle B(p_B)| H_{eff} (x) |\bar D(p_{\bar D}) X \rangle
        \langle \bar D(p_{\bar D}) X |
                H_{eff} (0) | B(p_B)\rangle 
\end{equation}
and make use of the fact that $m_c$ and $m_b$ are both large scales. 
We make these scales explicit by redefining the heavy quark fields 
in $H_{eff}$ by 
\begin{equation}
b (x) = b_v (x) e^{-im_b vx} \quad c (x) = c_{v'} (x) e^{-im_c v'x}
\end{equation}
where the velocities are defined as 
$p_B = m_B v$ and $p_D = m_D v'$.

Inserting this yields
\begin{align} \label{S}                                   
G (M^2) =& \sum_{X}\int d^4 x \, e^{-i(m_b v - m_c v')x} \\  
 &\qquad\qquad  \langle B(v) | \tilde{H}_{eff} (x)|\bar D(v') X \rangle 
\langle \bar D(v') X |  \tilde{H}_{eff} (0) | B(v)\rangle\,,
\nonumber
\end{align}
where $\tilde{H}_{eff}$ is obtained from $H_{eff}$ by the replacements
$b \to b_{v}$ and $c \to c_{v'}$.
Equation (\ref{S}) shows that the large momentum entering the game is 
$m_b v - m_c v'$. 

The next step is a short distance expansion of the matrix element 
appearing in (\ref{S}) yielding a power series in inverse powers 
of the large momentum 
$M=m_b v - m_c v'$
\begin{equation} \label{sde}
G(M^{2}) =\sum_{n = 0}^{\infty}\sum_{i}C^{(n)}_{i}(\mu)
\langle B(p_B) | \mathcal O^{(n)}_{i}  | B (p_B) \rangle |_\mu \,.
\end{equation}
The operators $\mathcal O^{(n)}_{i}$ depend on the 
final state $\bar D$--meson and are the analogue of the production 
operators as they appear in heavy quarkonia production \cite{BBL}
or in one-particle inclusive production in $e^+ e^-$ annihilation 
\cite{Mueller}. 
They are local and have the generic structure
\begin{equation}
 \mathcal O^{(n)}_{i}  = \sum_{X}[\bar c_{v'}\Gamma b_{v}]
|\bar D(v') X \rangle
\langle \bar D(v') X | [\bar b_{v}\Gamma' c_{v'}] \,,
\end{equation}
where $\Gamma^{(\prime)}$ denotes a combination of Dirac matrices
and covariant derivatives.

\begin{figure}
\begin{center}
\leavevmode
\epsfxsize=13.5cm
\epsffile[10 320 580 540]{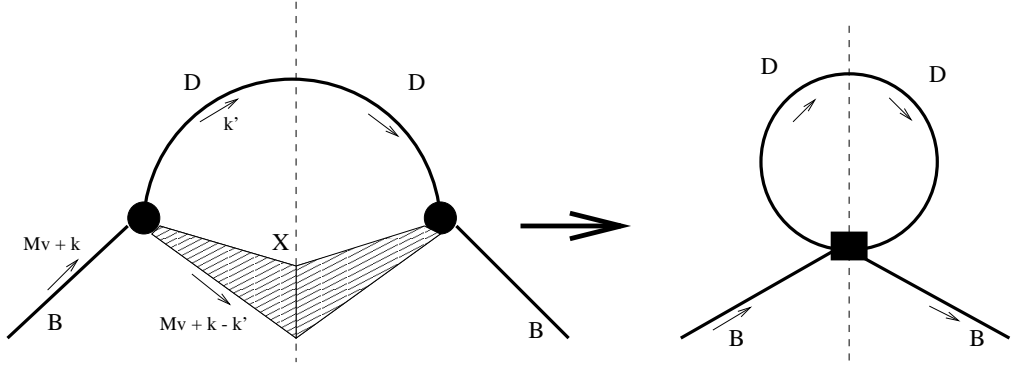}
\caption{Illustration of the short distance expansion of 
         \protect{(\ref{S})}.} 
\label{sdefig}
\end{center}
\end{figure}

The matrix elements of the  $\mathcal O^{(n)}_{i}$ between static $B$ 
meson states are universal functions of the velocity product $v \cdot v'$. We 
shall not give a detailed proof of factorization of the matrix elements 
into long and short distance contributions,  
rather we are aiming at a  phenomenological analysis of the one-particle 
inclusive semi-leptonic decays. We remark that 
the method is not as rigorous as in the case of fully inclusive decays, 
where the heavy mass expansion is derived by an operator product expansion.
However, fig.\ref{sdefig} makes the argument plausible. For $M^2$ large 
enough the large momentum flows through the state $| X \rangle$ and we
assume that parton-hadron duality holds for this part of the diagram, 
and hence we can compute this part in perturbative QCD. 

The dimension of $\mathcal O^{(n)}_{i}$ in (\ref{sde}) is $n + 6$ and hence 
$C_{i}^{(n)}/C_{i}^{(n+1)}$ is of the order $M$.
The leading term of the expansion involves dimension 6--operators
and we shall discuss in the present paper only this contribution. 
If we consider only this leading term, we may even replace the 
operators $b_v$ and $c_{c'}$ by static HQET quarks. In the following 
this replacement is understood.

The corrections to be expected can easily be estimated. 
Since the full momentum transfer is   
$Q = M + k$ where $k$ is the sum of the residual momenta
of the heavy $b$ and $c$ quark, the corrections to the leading term 
originate typically from 
\begin{equation}
Q^2 = M^2 + 2 M \cdot k + {\cal O}(\Lambda_{QCD}^2) 
    = M^2 \left(1 + \frac{2M \cdot k}{M^2} + \cdots \right) \, 
\end{equation}
and hence the corrections involve typically 
matrix elements of the form 
\begin{equation}
\sum_{X} \langle B(v) | [\bar{c}_{v'} \Gamma  b_v] 
|\bar D(v') X \rangle
\langle \bar D(v') X |
   (M \cdot iD)  [\bar{b}_v \Gamma^\dagger c_{v'}]  | B(v)\rangle
\end{equation}
where $D$ is the covariant derivative of QCD. This matrix element 
will be of order $M \cdot v \Lambda_{QCD}$ and 
consequently the method works as long as 
\begin{equation}
\frac{2M \cdot v}{M^2} \Lambda_{QCD} \ll 1 \,.
\end{equation}
$M^2$ and $M \cdot v$ are not independent variables, since they both 
can be expressed in terms of the velocity product $v \cdot v'$. Eliminating
$M \cdot v$ one obtains
\begin{equation}
\frac{\Lambda_{QCD}}{m_b} \left(\frac{m_b^2 - m_c^2}{M^2} + 1 \right)
\ll 1 \,,
\end{equation}
and hence it is obvious that the expansion breaks down for very small $M^2$. 
Here again a similar situation occurs as in the inclusive semi--leptonic 
decays, where the endpoint region may be described in terms of a shape 
function.

On the other hand one may ask wether the short distance expansion 
 works at all, and thus 
it is instructive to insert the maximal value for $M^2$ which is 
possible in a decay. One finds that the parameter 
$$
\frac{2 \Lambda_{QCD}}{m_b - m_c}  
$$ 
should be small compared to unity. Inserting the pessimistic value
$\Lambda_{QCD} = 500 $ MeV one finds that this parameter is about 
1/3, which justifies our approach for the spectra at least close 
to maximal $M^2$. 
In order to get the total rates an integration over the phase space of the 
$\bar D$ meson has to be performed. The details depend on the process
under consideration, but the typical size of the corrections can be 
estimated by computing some arbitrary phase space average.
Choosing a pase space measure as 
\begin{equation} 
d \bar \Phi  = 2m_{b}\,d\tilde p \, \frac{M^{2}}
{\sqrt{[(m_{c} + m_{b})^{2} - M^{2}][(m_{c} - m_{b})^{2} - M^{2} ]}}
\end{equation}
which yields very simple integrals we find
\begin{equation}
\langle \frac{M \cdot v}{M^2} \rangle  = 
\frac{\int d\bar\Phi  \frac{M \cdot v}{M^2}  }
{\int d\bar\Phi }  
=\frac{m_{b} \Lambda_{QCD}}{(m_b - m_c)^{2}} \approx \frac{1}{4}
\end{equation}
justifying the short distance expansion also for the rates.

The second type of corrections are the QCD radiative corrections 
which can be computed systematically. They will be of the order
$\alpha_s (M^2)$ and hence will be small enough to be treated 
perturbatively. As usual, the logarithms $\alpha_{s}(M^{2}) \ln(M^2)$ can be resummed by  
renormalization group methods; for the leading terms this will be done in 
section~\ref{sec:RG}.

In the present paper we study only the leading term of the expansion 
and focus on applications to weak interactions. In this case  we
need to consider a matrix element of a dimension-six operator 
involving the left handed currents. 

We consider the leptonic case in some detail; 
inserting the well known effective Hamiltonian for 
semi--leptonic decays we find
\begin{align}                                     
G& (M^2) = \frac{G_F^2}{2} |V_{cb}|^2P_{\mu \nu} (M) \\
& \sum_{X}  
\langle B(v) |[\bar{c}_{v'} \gamma^\mu (1-\gamma_5) b_v] 
|\bar D(v') X \rangle
\langle \bar D(v') X |                   
[\bar{b}_v \gamma^\nu (1-\gamma_5) c_{v'}]
                 | B(v)\rangle \,, \nonumber
\end{align}
where $P_{\mu \nu}$ is a tensor originating from contracting the lepton  
fields in the effective Hamiltonian. This tensor only depends on the vector
$M$ and hence has the form 
\begin{equation}
P_{\mu \nu} (M) = A(M^2) (M^2 g_{\mu \nu} - M_\mu M_\nu) 
                  +B(M^2) M_\mu M_\nu
\end{equation}
Neglecting the lepton masses, we obtain at tree level
\begin{equation}
A(M^2) = - \frac{1}{3 \pi} \Theta (M^2) \mbox{ and } B(M^2) = 0 \, .
\end{equation} 
Using
\begin{equation}
\bar{b}_v \fmslash{M} (1-\gamma_5) c_{v'} = 
(m_b - m_c) \bar{b}_v c_{v'} - (m_b + m_c) \bar{b}_v \gamma_5 c_{v'}
\end{equation}
we can write the leading order contribution as
\begin{eqnarray}
&& G (M^2) = \frac{G_F^2}{6 \pi} | V_{cb} |^2 4 m_B m_D 
\left[(m_B - m_D)^2 \eta_S (v \cdot v') \right. \\ 
&& \qquad \left.    + (m_B + m_D)^2 \eta_P (v \cdot v') 
                    - M^2 (\eta_V (v \cdot v') + \eta_A (v \cdot v')) \right]
\nonumber 
\end{eqnarray}    
where we have defined non-perturbative matrix elements as 
\begin{eqnarray} \label{etai}
4 m_B m_D \eta_S (v \cdot v')\!\!\!\! &=& \!\!\!\!
\sum_{X}      \langle B(v) | [\bar{c}_{v'} b_v] 
|\bar D(v') X \rangle
\langle \bar D(v') X |
[\bar{b}_v c_{v'}]
                               | B(v) \rangle  \\ \nonumber
- 4 m_B m_D \eta_P (v \cdot v')\!\!\!\! &=&\!\!\!\! \sum_{X} 
      \langle B(v) | [\bar{c}_{v'} \gamma_5 b_v]
|\bar D(v') X \rangle
\langle \bar D(v') X |                           
[\bar{b}_v \gamma_5 c_{v'}]
                         | B(v) \rangle  \\ \nonumber
4 m_B m_D \eta_V (v \cdot v')\!\!\!\! &=&\!\!\!\! \sum_{X} 
      \langle B(v) | [\bar{c}_{v'} \gamma^\mu b_v] 
|\bar D(v') X \rangle
\langle \bar D(v') X |                           
[\bar{b}_v \gamma_\mu c_{v'}]
                         | B(v)\rangle  \\ \nonumber
4 m_B m_D \eta_A (v \cdot v')\!\!\!\! &=&\!\!\!\! \sum_{X} 
      \langle B(v) | [\bar{c}_{v'} \gamma^\mu \gamma_5 b_v] 
|\bar D(v') X \rangle
\langle \bar D(v') X |                           
[\bar{b}_v \gamma_\mu \gamma_5 c_{v'}]
                         | B(v)\rangle  .
\end{eqnarray}
Once radiative corrections are taken into account, these operators mix 
with the corresponding operators where the $b$ and the $c$ quark are coupled
to a color octett: 
\begin{eqnarray} \label{rhoi}
4 m_B m_D \rho_S (v \cdot v')\!\!\!\! &=&\!\!\!\! \sum_{X} 
      \langle B(v)  |[\bar{c}_{v'} T^a b_v] 
|\bar D(v') X \rangle
\langle \bar D(v') X |
                          [\bar{b}_v T^a c_{v'}]
                         | B(v)  \rangle  \\ \nonumber 
- 4 m_B m_D \rho_P (v \cdot v')\!\!\!\! &=&\!\!\!\! \sum_{X} 
      \langle B(v)  |[\bar{c}_{v'}  \gamma_5 T^a b_v] 
|\bar D(v') X \rangle
\langle \bar D(v') X |
                          [\bar{b}_v \gamma_5 T^a c_{v'}]
                         | B(v)  \rangle  \\ \nonumber
4 m_B m_D \rho_V (v \cdot v')\!\!\!\! &=&\!\!\!\! \sum_{X} 
      \langle B(v)|[\bar{c}_{v'}  \gamma^\mu T^a b_v] 
|\bar D(v') X \rangle
\langle \bar D(v') X |
                         [\bar{b}_v \gamma_\mu T^a c_{v'}]
                         | B(v)\rangle  \\ \nonumber
4 m_B m_D \rho_A (v \cdot v')\!\!\!\! &=& \nonumber \\
&&\hspace{-1cm}\sum_{X} 
      \langle B(v) |[\bar{c}_{v'} \gamma^\mu \gamma_5 T^a b_v] 
|\bar D(v') X \rangle
\langle \bar D(v') X |
                          [\bar{b}_v \gamma_\mu \gamma_5 T^a c_{v'}]
                         | B(v) \rangle  \nonumber
\end{eqnarray}
Note that we are using parton-hadron duality and thus $| X \rangle$ is 
expressed in terms of QCD degrees of freedom and thus the matrix elements
$\rho_i$ are nonvanishing.

\section{Renormalization Group Improvement}
\label{sec:RG}
Under renormalization the matrix elements
(\ref{etai},\ref{rhoi}) become scale dependend quantities 
$\eta_{i}(v \cdot v',\mu)$, $\rho_{i}(v \cdot v',\mu)$.
We chose to construct the short distance expansion  at
an intermediate scale $\bar m = 1/2(m_{b} + m_{c})$, where
both the $b$-- and $c$--Quark are treated as static fields
described by the HQET. However, the typical scale of the hadronic matrix
elements is a low hadronic scale $\mu = \Lambda$. 
Using the renormalization group the matrix elements can be scaled from 
the matching scale $\bar m$ down to the low scale $\Lambda$.
In our numerical analysis we use $\alpha_{s}(\Lambda)=1$,
where $\alpha_{s}(\mu)$ is the one loop expression for the running
coupling constant.

To this end one has to renormalize
the operators
\footnote{Note that at least to one loop order 
the renormalization properties of these operators
are identical to those of the operators
\begin{equation}
\mathcal O^{(1)}_{i} = [\bar c_{v'}\Gamma_{i}b_{v}] [\bar b_{v}\Gamma_{i}c_{v'}]
\qquad
\mathcal O^{(8)}_{i} = [\bar c_{v'}\Gamma_{i}T^{a}b_{v}] 
[\bar b_{v}\Gamma_{i}T^{a}c_{v'}]\,.
\end{equation}
since the UV behaviour is independent of the states, including the 
$\bar{D}$ appearing in the final state.} 
%
%This results from the fact
%that the UV--divergencies do not depend on  
%a $\bar D$--meson in the 
%final or a $D$--meson in the initial state as long as we are calculating
%the contribution to a forward matrixelement of a hermitian operator 
%which has to be a real
%quantity. Otherwise one is faced with imaginary parts when crossing 
%the $\bar D$--meson into the initial state.}
\begin{align}
\mathcal O^{(1)}_{i} =& \sum_{X}[\bar c_{v'}\Gamma_{i}b_{v}] 
|\bar D(v') X \rangle
\langle \bar D(v') X |
                       [\bar b_{v}\Gamma_{i}c_{v'}]
\nonumber \\
\mathcal O^{(8)}_{i} =& \sum_{X}[\bar c_{v'}\Gamma_{i}T^{a}b_{v}] 
|\bar D(v') X \rangle
\langle \bar D(v') X |
                       [\bar b_{v}\Gamma_{i}T^{a}c_{v'}]
\end{align}
where $\Gamma_{i} = 1, \gamma_{5}, \gamma_{\mu},
\gamma_{\mu}\gamma_{5}$. Because of heavy quark spin symmetry 
mixing occures only between singlett and octett operators
corresponding to one specific Dirac structure $\Gamma_{i}$.
That means a basis closing under renormalization is given by
$\mathcal O^{(1)}_{i}$ and $\mathcal O^{(8)}_{i}$
for every individual $i$.

\begin{figure}
\begin{center}
\leavevmode
\epsfxsize=9cm
\epsffile[100 450 490 750]{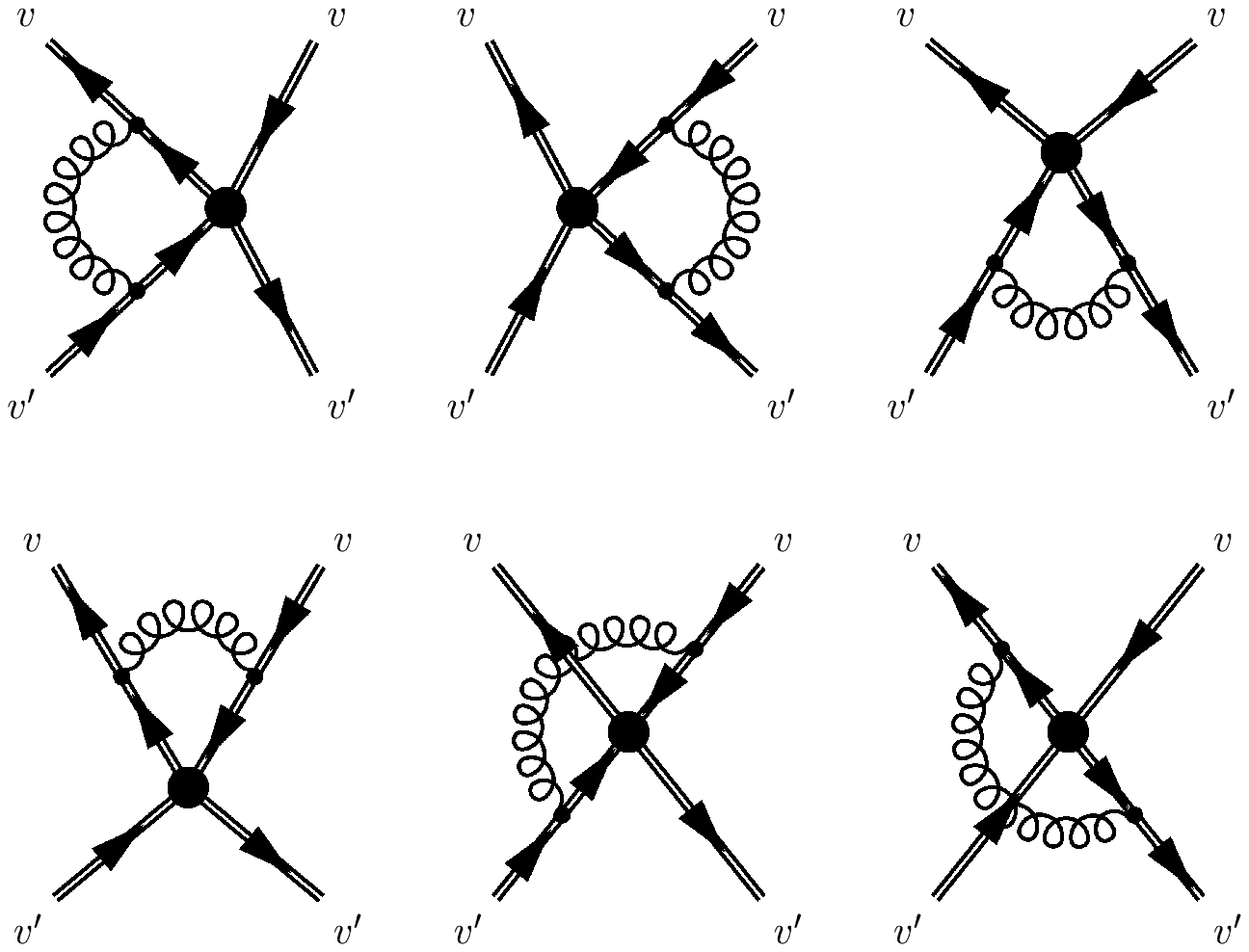}
\caption{\label{fig:diagrams} Feynman diagrams contributing to the 
one loop anomalous dimensions. The blob represents generically the operators
corresponding to the $\eta_i$ and $\rho_i$.}
\end{center}
\end{figure}

The mixing properties of the operators translate into that of their
matrix elements. Therefore we can formulate the renormalization
group equation directly in terms 
of the $\eta_{i}(v \cdot v',\mu)$ and  $\rho_{i}(v \cdot v',\mu)$
as follows:
\begin{equation}
\label{eq:RGeqation}
\begin{aligned}
\frac{d}{d\ln \mu}\eta_{i}(v \cdot v',\mu) &= \gamma_{11}\eta_{i}(v \cdot v',\mu)
                                       + \gamma_{18}\rho_{i}(v \cdot v',\mu)\\
\frac{d}{d\ln \mu}\rho_{i}(v \cdot v',\mu) &= \gamma_{81}\eta_{i}(v \cdot v',\mu)
                                       + \gamma_{88}\rho_{i}(v \cdot v',\mu)
\end{aligned}
\end{equation}
Since we restrict ourselves to the leading logarithmic approximation,
it suffices to know the one loop anomalous dimensions. These
are given by the divergent 
parts of the Feynman diagrams shown in figure \ref{fig:diagrams}
supplemented by wave function renormalization
of the heavy quark fields
\begin{equation}
\begin{aligned}
\gamma_{11} &= (\frac{\alpha_{s}}{\pi})(N_{c}-\frac{1}{N_{c}}) K(v \cdot v') &
\qquad\gamma_{18} &= -(\frac{\alpha_{s}}{\pi})2K(v \cdot v') \\
\gamma_{81} &=(\frac{\alpha_{s}}{\pi}) \frac{1}{2}(\frac{1}{N_{c}^{2}} - 1) K(v \cdot v') &
\qquad\gamma_{88} &=(\frac{\alpha_{s}}{\pi}) \frac{1}{N_{c}}K(v \cdot v') 
\end{aligned}
\end{equation}
where $N_{c} = 3$ is the number of colors and 
\begin{equation}
K(v \cdot v') = 1 - v \cdot v' \mathrm{Re}[r(v \cdot v')] \qquad r(z) = \frac{\ln(z + \sqrt{z^{2} - 1})}
                                            {\sqrt{z^{2} - 1}}\,.
\end{equation}
The function $r(v \cdot v')$ typically appears in the anomalous dimensions 
of velocity chanching 
heavy quark currents \cite{HQET}. 

In our case only the real part of $r(v \cdot v')$ shows up  in 
the anomalous dimensions, since the corresponding
Feynman amplitudes contribute to the forward matrix element
of a hermitian operator which has to be real. 
Note that individual Feynman
diagrams develop imaginary parts which drop out in the sum.     

Solving (\ref{eq:RGeqation}) we express the matrix
elements $\eta_{i}, \rho_{i}$ at the scale $\bar m$ 
in terms of their value at an arbitrary
scale $\mu$:
\begin{align}
E_{i}(v \cdot v') &= \eta_{i}(v \cdot v',\bar m) =
C_{11}(v \cdot v',\mu) \eta_{i}(v \cdot v',\mu) +
C_{18}(v \cdot v',\mu) \rho_{i}(v \cdot v',\mu) \nonumber\\
R_{i}(v \cdot v') &= \rho_{i}(v \cdot v',\bar m) =
C_{81}(v \cdot v',\mu) \eta_{i}(v \cdot v',\mu) +
C_{88}(v \cdot v',\mu) \rho_{i}(v \cdot v',\mu) 
\end{align}
The coefficient functions $C_{ij}(v \cdot v',\mu)$ are given by
\begin{equation}
\label{C1C2}
\begin{aligned}
C_{11}(v \cdot v',\mu) &=\frac{1}{N_{c}^{2}} + (1 -
\frac{1}{N_{c}^{2}})\zeta(v \cdot v',\mu)  \\
C_{18}(v \cdot v',\mu) &=\frac{2}{N_{c}}\biggl(1 - \zeta(v \cdot v',\mu)\biggr)  \\
C_{81}(v \cdot v',\mu) &=\frac{1}{2N_{c}}(\frac{1}{N_{c}^{2}}-1)\biggl(\zeta(v \cdot v',\mu)-1\biggr)
 \\
C_{88}(v \cdot v',\mu) &=1+ \frac{1}{N_{c}^{2}}\biggl(\zeta(v \cdot v',\mu)-1\biggr) 
\end{aligned}
\end{equation} 
where 
\begin{displaymath}
\zeta(v \cdot v',\mu) = \biggl(
\frac{\alpha_{s}(\mu)}{\alpha_{s}(\bar m)}
\biggr)^{\frac{N_{c}}{2\beta_{0}} K(v \cdot v')}
\end{displaymath}
with $\beta_{0} = (33-2N_{f})/12$ and $N_f=3$ for three active quark flavours.

Note that in the case of semi--leptonic decays 
only the functions $E_{i}(v \cdot v')$ are needed,
since in LLA there are no octett contributions at the matching scale.

\section{One-particle Inclusive Semi--leptonic Decays} 
We shall first try to understand the data on the decays 
$B \to \bar D X\ell^+ \nu$. In order to do this we need to have some 
idea about the matrix elements $\eta_i$ and $\rho_i$ ($i=S,P,V,A$)
which are defined in (\ref{etai}) and (\ref{rhoi}). 
We shall work to leading order in the $1/M$ expansion and 
hence identify $m_c = m_{D} = m_{D^*}$ and $m_{b} = m_{B}$. 

We are aiming at the energy spectrum of the $\bar D$ meson in the 
one-particle inclusive decays of the type $B \to \bar DX\ell^+ \nu$.
The rate is obtained by 
integrating over the phase space of the $\bar D$. 
Taking into account renormalization one gets
\begin{eqnarray} \label{master}
 \frac{d\Gamma}{dy} &&= \frac{1}{2m_B} G(M^2) \frac{m_{D}^2}{4 \pi^2} 
                     \sqrt{y^2 -1} \\ \nonumber 
&& = \frac{G_F^2}{12 \pi^3} | V_{cb} |^2 m_{D}^3   \sqrt{y^2 -1}
\left[(m_B - m_{D})^2 E_S (y) \right. \\ 
&& \nonumber \qquad \left.    + (m_B + m_{D})^2 E_P (y) 
                    - M^2 (E_V (y) + E_A (y)) \right] 
\end{eqnarray}    
where $y = v \cdot v'$ and 
the $E_i$ ($i=S,P,V,A$) are 
the renormalization group invariant combinations of the 
$\eta_i$ and $\rho_i$ 
\begin{equation}
E_i (v \cdot v') = C_{11}(v \cdot v',\mu) \eta_i (v \cdot v',\mu) + C_{18}(v \cdot v',\mu) \rho_i(v \cdot v',\mu)
\,.
\end{equation}
The Wilson coefficients $C_{11}$ and $C_{18}$ have been given in 
(\ref{C1C2}). 

To get some expression for the functions  $\eta_i$ and $\rho_i$ ($i=S,P,V,A$)
we first observe that at $v \cdot v' =1$ the inclusive rate is 
saturated by the exclusive decays into the lowest-lying spin 
symmetry doublet $\bar D$ and $\bar D^*$. Furthermore, at this point only 
the $\eta_i$ contribute, since  $C_{18} (v \cdot v' = 1) = 0$. 
The $\bar D^*$ subsequently decays into
$\bar D$ mesons and thus at $v \cdot v' = 1$ the sum of the exclusive rates for 
$B \to \bar D \ell^+ \nu$ and $B \to \bar D^* \ell^+ \nu$ 
is equal to the one-particle inclusive semi--leptonic rate
$B \to \bar D \ell^+ \nu X$, which is again 
equal to the fully inclusive rate $B \to X_{\bar c} \ell^+ \nu$. 
In other words, at this point there are no decays into other charmed 
hadrons than $\bar D$ mesons. 

Off this point things become more complicated. However, as far as the 
total rates are concerned, still the exclusive decays 
$B \to \bar D \ell^+ \nu$ and $B \to \bar D^* \ell^+ \nu$ saturate 
the fully inclusive rate $B \to X_{\bar c} \ell^+ \nu$ at a level of 
about 70\%. Since the $\bar D^*$ decay all into $\bar D$ mesons, it is 
certainly a good starting point to approximate the $\eta_i$ by 
something one obtains from the sum of the exclusive decays. 
In other words, we shall express the $\eta_i$ in terms of the 
Isgur-Wise function \cite{IsgurWise}.

The approximation we are going to use corresponds to some kind of 
factorization assumption formulated for $G(M^{2})$.
The functions $\eta_{i}$ are defined by the 
matrix elements (\ref{etai}) and we shall approximate these 
matrix elements.
However, as with the usual factorization, our approximation 
is not a scale invariant concept, and hence we have to define, 
at which scale it should hold. 
At a small hadronic scale $\Lambda$ we replace in (\ref{etai}) 
\begin{eqnarray} \label{fact0}
&& \sum_{X}\langle B(v) | [\bar{c}_{v'} \Gamma_i b_v] 
|\bar D(v') X \rangle
\langle \bar D(v') X |
                        [\bar{b}_v \Gamma_i c_{v'}]
                      | B(v) \rangle|_{\mu=\Lambda} 
\longrightarrow  \\ && \nonumber 
%\hspace{1.3cm}   
\langle B(v) |  \bar{c}_{v'} \Gamma_i b_v | \bar D(v') \rangle|_{\mu=\Lambda}
\langle \bar D(v')| \bar{b}_v \Gamma_i c_{v'} | B(v) \rangle|_{\mu=\Lambda} \\ 
\nonumber && + \sum_{Y(\bar D^*)}  
\langle B(v) |  \bar{c}_{v'} \Gamma_i b_v
                         |\bar D(v')  Y(\bar D^*) \rangle|_{\mu=\Lambda} 
\langle \bar D(v') Y(\bar D^*) | \bar{b}_v \Gamma_i c_{v'}  | B(v) \rangle|_{\mu=\Lambda}
\end{eqnarray}
where $Y(\bar D^*)$ is defined by $\bar D^* \to \bar D Y(\bar D^*)$, i.e.\
$Y(\bar D^*)$ is 
either a pion or a photon originating from a $\bar D^*$ decay. 
In the following we shall call this replacement factorization, since it 
is closely related to the factorization assumption known from 
non-leptonic decays.  We get, 
again schematically
\begin{eqnarray} \label{fact1}
&& \sum_{X}\langle B(v) |[\bar{c}_{v'} \Gamma_i b_v] 
|\bar D(v') X \rangle
\langle \bar D(v') X |
                       [\bar{b}_v \Gamma_i c_{v'}]
                      | B(v) \rangle|_{\mu=\Lambda}
\longrightarrow  \\ &&  \nonumber 
%\hspace{1cm}   
\langle B(v) | \bar{c}_{v'} \Gamma_i b_{v} | \bar D(v') \rangle|_{\mu=\Lambda} 
\langle \bar D(v')| \bar{b}_{v} \Gamma_i c_{v'} | B(v) \rangle|_{\mu=\Lambda} \\ 
\nonumber && + \sum_{Pol} 
\langle B(v) | \bar{c}_{v'} \Gamma_i b_{v} |\bar D^*(v',\epsilon) \rangle|_{\mu=\Lambda} 
\langle \bar D^*(v',\epsilon) | \bar{b}_{v} \Gamma_i c_{v'} | B(v) 
                           \rangle|_{\mu=\Lambda} \nonumber \\
&&\qquad \cdot \mathrm{Br}(\bar D^* \to \bar D Y(\bar D^*))  \nonumber
\end{eqnarray}
where the sum runs over the polarization states of the $D^*$. 
In (\ref{fact1}) we have used the narrow width approximation for
the $\bar D^{*}$ in the intermediate state.

The matrix elements appearing in the factorized expression
(\ref{fact1}) can all be expressed 
in terms of the Isgur--Wise function:
\begin{align}
\langle B(v) | \bar{c}_{v'} \Gamma_i b_{v} | \bar D(v') \rangle|_{\mu} 
 &= \frac{1}{4}\sqrt{m_{B}m_{D}}\mathrm{Tr}
\{\gamma_{5}(1 + \fmslash{v}) \Gamma_{i}(1 + \fmslash{v'})\gamma_{5} \}
\xi(v \cdot v',\mu) \nonumber\\
\langle B(v) | \bar{c}_{v'} \Gamma_i b_{v} | \bar D^{*}(v',\epsilon) \rangle|_{\mu} 
 &= \frac{1}{4}\sqrt{m_{B}m_{D^{*}}}\mathrm{Tr}
\{\gamma_{5}(1 + \fmslash{v}) \Gamma_{i}(1 + \fmslash{v'})\fmslash{\epsilon} \}
\xi(v \cdot v',\mu) \label{IGW}
\end{align}
From this we get
\begin{equation}
\eta_{i}(v \cdot v',\mu) = \frac{|X(v \cdot v')|^{2}}{ C^{2}_3 (v \cdot v',\mu)}
\biggl[c_{i}(v \cdot v') + c^{*}_{i}(v \cdot v')\mathrm{Br}(\bar D^* \to \bar D X(\bar D^*))
\biggr]
\end{equation}
where
\begin{align}
c_{i}(v \cdot v') &=\frac{1}{16}
|\mathrm{Tr}
\{\gamma_{5}(1 + \fmslash{v}) \Gamma_{i}(1 + \fmslash{v'})\gamma_{5} \}|^{2}
\nonumber \\
c^{*}_{i}(v \cdot v') &=\frac{1}{16} \sum_{Pol}
|\mathrm{Tr}
\{\gamma_{5}(1 + \fmslash{v}) \Gamma_{i}(1 + \fmslash{v'})\fmslash{\epsilon}
\}|^{2} \nonumber
\end{align}
and $X(v \cdot v')$ is the renormalization group invariant combination 
\begin{equation}
X(v \cdot v') = C_3 (v \cdot v',\mu) \xi(v \cdot v',\mu) \,.
\end{equation}
The Wilson coefficient $C_3 (v \cdot v',\mu)$ renormalizing the 
$b_{v} \to c_{v'}$ current is known to two loops, 
but since we computed 
$C_{11}$ and $C_{18}$ only to one loop, it is sufficient to insert the one 
loop result
\begin{equation}
C_3 (v \cdot v',\mu) = \biggl(
\frac{\alpha_s(\mu)}{\alpha_s(\bar m)}
\biggr)^{\frac{1}{2\beta_0}\gamma_{hh}(v \cdot v')} 
\end{equation}
where
\begin{equation}
\gamma_{hh}(v \cdot v') = \frac{1}{2}(N_{c} - \frac{1}{N_{c}})\biggl(1 - v
\cdot v' \, r(v \cdot v') 
\biggr) \,.
\end{equation}

The factorization assumption yields expressions for the matrix elements 
$\eta_i$ at the small scale $\Lambda$, but it does not tell us 
anything about the color octett contributions $\rho_i$.
It is well known that factorization should hold
in the limit $N_{c} \to \infty$. This fact is indeed reflected in the 
$N_{c}$--dependence of the Wilson coefficients, since
\begin{align}
\lim_{N_{c} \to \infty} C_{18} &=\lim_{N_{c} \to \infty} C_{81} = 0 \nonumber
\\
\lim_{N_{c} \to \infty} C_{88} &= 1 \\ 
\lim_{N_{c} \to \infty} C_{11} &= |C_{3}|^{2}  \nonumber
\end{align}
and thus the dimension 6--operators renormalize as products 
of dimension 3--currents and factorization 
becomes scale independent.
This does still not tell us much about the $\rho_{i}$, but a natural
assumption is that they are of the order $1/N_C$ and hence we 
shall take $\rho_{i}$ to be constant with
$\rho_i (v \cdot v',\mu) = 1/N_C$. This simple ansatz, 
ignoring a possible dependence on $v \cdot v'$, does not 
introduce large uncertainties for the one-particle inclusive semi--leptonic decays, 
since the $\rho_i$ are only induced through radiative corrections.   

In the following we shall consider the decays of the $B^+$ and the $B^0$, 
both of which contain a $\bar{b}$ quark undergoing a semi--leptonic transition
$\bar{b} \to \bar{c} \ell^+ \nu $. In the heavy mass limit for the 
$c$ quark the final states involving a $c$ quark ($D^0$ or $D^+$ states) 
are suppressed, since this would involve a $c\bar{c}$ pair creation. To 
leading order in $1/m_c$ the possible decays are thus
\begin{eqnarray} \label{sinclmodes}
& B^+ \to \overline{D}^0 \ell^+ \nu X \qquad 
& B^+ \to D^- \ell^+ \nu X \\
& B^0 \to \overline{D}^0 \ell^+ \nu X \qquad 
& B^0 \to D^- \ell^+ \nu X \nonumber \,.
\end{eqnarray}
Since many of the $\bar D$ mesons originate from $\bar D^*$ decays we 
have to take into account the relevant  branching ratios of the 
$\bar D^*$ mesons into the $\bar D$ mesons of different charge.  
We use  \cite{PDG}
\begin{eqnarray} \label{Dbrs}
&& \mbox{Br} (D^{*-} \to \overline{D}^0 X) \approx 68\% \quad
\mbox{Br} (D^{*-} \to D^- X) \approx 32\% \\
&& \mbox{Br} (\bar D^{*0} \to \bar D^0 X) \approx 100\% 
\nonumber
\end{eqnarray}
and hence we can have the following decay chains 
\begin{eqnarray} 
& B^+ \to \overline{D}^{*0} \ell^+ \nu \qquad
\stackrel{\mbox{Br} (\bar D^{*0} \to \bar D^0 X)}{\longrightarrow} 
\qquad B^{+} \to \overline{D}^0 \ell^+ \nu X \nonumber \\
& B^0 \to D^{*-}  \ell^+ \nu  \qquad
\stackrel{\mbox{Br} (D^{*-} \to D^- X)}{\longrightarrow} 
\qquad B^0 \to D^-  \ell^+ \nu X  \label{chains} \\
& B^0 \to D^{*-}  \ell^+ \nu  \qquad
\stackrel{\mbox{Br} (D^{*-} \to \overline{D}^0 X)}{\longrightarrow} 
\qquad B^0 \to \overline{D}^0 \ell^+ \nu X \nonumber\,,
\end{eqnarray} 
where the arrow indicates that -- in addition to the
direct decay channel $B \to \bar D \ell^+ \nu$ -- 
the exclusive mode on the l.h.s. contributes 
to the one-particle inclusive rate on the 
r.h.s. weighted with the branching ratios (\ref{Dbrs}).

We shall label the  $E_i$ for the different decay modes (\ref{sinclmodes}) 
with a superscript indicating the initial $B$ and the final $\bar D$ meson.
Taking into account the $\bar D^*$ branching ratios (\ref{Dbrs}) we arrive 
at the following expressions for the $E_i$  involved in the $B^+$ decays:
\vspace{.5cm}
\begin{eqnarray}
E^{B^+\overline{D}^0}_S (y) &=&\nonumber \frac{C_{11} (y,\Lambda)}{C_3^2 (y,\Lambda)} 
              \frac{1}{4} (y+1)^2 |X(y)|^2 + C_{18} (y,\Lambda) \frac{1}{N_C} 
\\  \nonumber 
E^{B^+\overline{D}^0}_P (y) &=& \frac{C_{11} (y,\Lambda)}{C_3^2 (y,\Lambda)}
              \frac{1}{4} (y^2-1) |X(y)|^2 + C_{18} (y,\Lambda) \frac{1}{N_C}
\\ \nonumber 
E^{B^+\overline{D}^0}_V (y) &=& \frac{C_{11} (y,\Lambda)}{C_3^2 (y,\Lambda)}
              \frac{1}{2} y (y+1) |X(y)|^2 + C_{18} (y,\Lambda) \frac{1}{N_C}
\\ 
E^{B^+\overline{D}^0}_A (y) &=& - \frac{C_{11} (y,\Lambda)}{C_3^2 (y,\Lambda)}
            \frac{1}{2} (y+2) (y+1) |X(y)|^2 + C_{18} (y,\Lambda) \frac{1}{N_C}
\end{eqnarray}
%\vspace{.5cm}
In the $B^0$ decays we have to take into account the $D^{*-}$ 
branching ratios as
\vspace{.5cm}
\begin{eqnarray}
E^{B^0\overline{D}^0}_S (y) &=& \nonumber C_{18} (y,\Lambda) \frac{1}{N_C} 
\\ \nonumber 
E^{B^0\overline{D}^0}_P (y) &=& \frac{C_{11} (y,\Lambda)}{C_3^2 (y,\Lambda)}
       Br(D^{*-} \to \overline{D}^0 X) \frac{1}{4} (y^2 - 1) |X(y)|^2 
              + C_{18} (y,\Lambda) \frac{1}{N_C}
\\ \nonumber 
E^{B^0\overline{D}^0}_V (y) &=& \frac{C_{11} (y,\Lambda)}{C_3^2 (y,\Lambda)}
       Br(D^{*-} \to \overline{D}^0 X) \frac{1}{2} (y^2 - 1) |X(y)|^2 
              + C_{18} (y,\Lambda) \frac{1}{N_C}
\\  \nonumber 
E^{B^0\overline{D}^0}_A (y) &=& - \frac{C_{11} (y,\Lambda)}{C_3^2 (y,\Lambda)}
       Br(D^{*-} \to \overline{D}^0 X) \frac{1}{2} (y+2) (y+1) |X(y)|^2 
\\  && \qquad + C_{18} (y,\Lambda) \frac{1}{N_C}
\end{eqnarray}
and 
\begin{eqnarray}\label{B0D0}
E^{B^0D^-}_S (y) &=& \frac{C_{11} (y,\Lambda)}{C_3^2 (y,\Lambda)} 
                     \frac{1}{4}(y+1)^2 |X(y)|^2   
                     + C_{18} (y,\Lambda) \frac{1}{N_C} 
\\ \nonumber 
E^{B^0 D^-}_P (y) &=& \frac{C_{11} (y,\Lambda)}{C_3^2 (y,\Lambda)}
       Br(D^{*-} \to D^- X) \frac{1}{4} (y^2 - 1) |X(y)|^2 
              + C_{18} (y,\Lambda) \frac{1}{N_C}
\\ \nonumber 
E^{B^0 D^-}_V (y) &=& \frac{C_{11} (y,\Lambda)}{C_3^2 (y,\Lambda)} 
\left( \frac{1}{2}(y+1) + Br(D^{*-} \to D^- X) 
       \frac{1}{2} (y^2 - 1) \right) |X(y)|^2 
\\ \nonumber && \qquad   + C_{18} (y,\Lambda) \frac{1}{N_C}
\\  \nonumber 
E^{B^0 D^-}_A (y) &=& - \frac{C_{11} (y,\Lambda)}{C_3^2 (y,\Lambda)} 
       Br(D^{*-} \to D^- X) \frac{1}{2} (y+2) (y+1) |X(y)|^2 
\\ \nonumber && \qquad + C_{18} (y,\Lambda) \frac{1}{N_C} \,.
\end{eqnarray}
Note that at $v \cdot v' = 1$ we have simply the sum of the exclusive 
channels $B \to \bar D \ell^+ \nu$ and $B \to \bar D^* \ell^+ \nu$, 
where the $\bar D^*$ component is weighted with the appropriate $\bar D^*$ 
branching ratios, since here $C_{11} = C_3 =1$ and $C_{18} = 0$. 
Off the point $v \cdot v' = 1$ we still have 
$ C_{11} / (C_3)^2 \approx 1$ but there is also an 
additional contribution from the octett contributions $\rho_i$. 
As we shall see, this additional contributions are consistent with 
the data, despite of our crude estimate. 

Inserting these lenghty expressions into the master formula 
(\ref{master}) one obtains expressions for the one-particle inclusive 
energy spectra of the $\bar D$ mesons. To leading order in $1/m_c$ the energy 
of the $\bar D^*$ meson is equal to the energy of the $\bar D$ meson originating 
from the decay $\bar D^* \to \bar DX$ since $X$ is soft of the order $1/m_c$. 

In order to actually obtain numbers one needs the Isgur Wise function as 
an input. A good fit to the experimental data is obtained already with 
a linear function, which is fitted to the renormalization group invariant
$X(y)$
\begin{equation}
X(y) = 1 - a (y-1) \quad \mbox{ with } \quad a = 0.84 \quad \mbox{\cite{CLEO}} \,.
\end{equation}

\begin{figure}
\begin{center}
\leavevmode
\epsfysize=10cm 
\epsffile{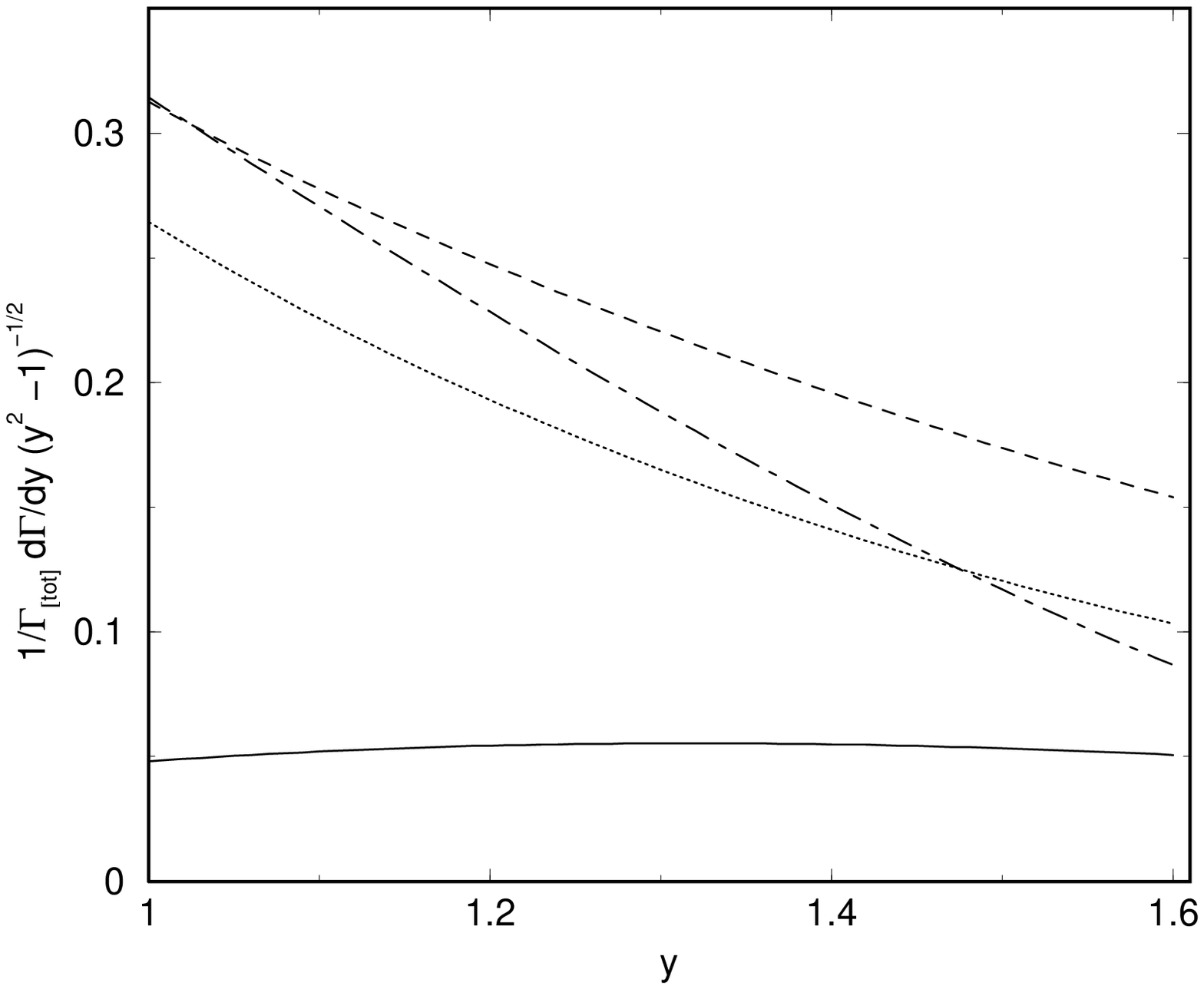} 
\end{center}
\caption{Decay spectra of the one-particle inclusive decays. 
         Solid line: $B \to D^- \ell^+ \nu X$, 
         dotted line: $B \to \overline{D}^0 \ell^+ \nu X$,
         dashed line: $B \to (D^- + \overline{D}^0) \ell^+ \nu X$,
         dashed-dotted line: $B^0 \to (D^- + D^{*-}) \ell^+ \nu $.}
\label{XC}
\end{figure}

In figure \ref{XC} we plot the spectra of the $D$ meson 
for the combined rates 
\begin{eqnarray*}
\frac{d \Gamma}{dy}(B \to D^- \ell^+ \nu X)    
&=& \frac{1}{2}\left(\frac{d \Gamma}{dy}(B^+ \to D^- \ell^+ \nu X) 
  + \frac{d \Gamma}{dy}(B^0 \to D^- \ell^+ \nu X) \right) \\
\frac{d \Gamma}{dy}(B \to \overline{D}^0 \ell^+ \nu X) 
&=& \frac{1}{2}\left( 
  \frac{d \Gamma}{dy}(B^+ \to \overline{D}^0 \ell^+ \nu X) 
  + \frac{d \Gamma}{dy}(B^0 \to \overline{D}^0 \ell^+ \nu X)  
    \right)
\end{eqnarray*}
and compare it to the sum of the exclusive decays 
$B^0 \to D^- \ell^+ \nu$ 
and $B^0 \to D^{*-} \ell^+ \nu$ .  

One may also integrate the spectra to obtain a total rate for the  
one-particle inclusive semi--leptonic processes. In table~\ref{XTC} we compare the
rates we obtain from our approach with the experimental data from 
\cite{PDG}.

\begin{table}
\begin{center}
\begin{tabular}{|l|l|l|}
\hline
Mode & Br (theory) & Br (data from \protect{\cite{PDG}}) \\
\hline 
$B \to D^- \ell^+ \nu X$              & $2.3\%$ & $(2.7 \pm 0.8)\%$ \\
$B \to \overline{D}^0 \ell^+ \nu X$   & $6.9\%$ & $(7.0 \pm 1.4)\%$ \\
$B^0 \to D^- \ell^+ \nu $            &     & $(1.5 \pm 0.5)\%$ \\
$B^0 \to D^{-*} \ell^+ \nu $         &     & $(4.68\pm 0.25)\%$\\
$B \to \bar D^{**} \ell \nu$              & $(3.5 \pm 0.6)\%$ & $(2.7 \pm
0.7)\%$     \\
$B \to \mbox{non-}\bar D \ell^+ \nu$      & $(0.6 \pm 0.4)\%$ &     \\
\hline
\end{tabular}
\end{center}
\caption{Comparison of our results with data. To get branching ratios, 
         we used $\tau_{B^+} = \tau_{B^0} = 1.55 $ ps. Here $\bar D^{**}$ 
         denotes any final state with a $\bar D$ meson which does not come 
         from the exclusive decays listed in row three and four. The last 
         two rows are commented in the text.}
\label{XTC}
\end{table}    

Table~\ref{XTC} and also figure \ref{XC} exhibit 
a few interesting features. First of all the 
experimental data are well reproduced. Furthermore, although we have 
used the assumption (\ref{fact1}) our result is not simply the sum of the 
inclusive decays $B \to \bar D \ell^+ \nu$ and $B \to \bar D^* \ell^+ \nu$,
since (\ref{fact1}) is a scale dependent statement. We assume that 
(\ref{fact1}) holds at the small scale $\Lambda$; running up to 
the matching scale $\overline{m}$  yields a significant 
contribution from gluon exchanges. We interpret these contributions as
$B \to \bar D^{**} \ell^+ \nu$ where $\bar D^{**}$ now stands for all 
$\bar D$-meson final states, which do not originate from 
$B \to \bar D \ell^+ \nu$ or $B \to \bar D^* \ell^+ \nu$. Although the 
ansatz for the octett matrix elements $\rho_i$ is extremely simple, we
obtain a reasonable number, namely  
$
\mbox{Br}(B \to \bar D^{**} \ell^+ \nu) \approx 32\% \times  
\mbox{Br}(B \to X_{\bar c} \ell^+ \nu)
$
where we use $Br(B \to X_{\bar{c}} \ell^+ \nu) = (10.4 \pm 0.4)\%$ from 
\cite{PDG}. 
From figure \ref{XC} it is obvious that the $\bar D^{**}$--contribution
vanishes at $v \cdot v' = 1$ as required by the heavy quark limit.

The last row of table~\ref{XTC} gives the branching ratio for decays 
which do not have a $\bar D$ meson in the final state, rather some other 
charmed hadron. The only other ground state hadron is a $\bar \Lambda_{c}$ 
so this should be the branching ratio for 
$B \to \bar \Lambda_{c}  X \ell^+ \nu$ for which we obtain 
$\mbox{Br} (B \to \bar \Lambda_{c}  X \ell^+ \nu) = 6\% \times  
\mbox{Br}(B \to X_{\bar c} \ell^+ \nu)$. This is what one would expect 
on the basis of the naive reasoning that a heavy quark hadronizes 
into a baryon with a branching ratio of about ten percent. 

\section{Conclusions}
Exclusive semi--leptonic as well as fully inclusive decays of 
heavy hadrons have a 
well established basis in QCD. While in the former case it is the 
heavy mass limit of QCD, formulated as an effective theory (HQET), 
in the latter case it is the heavy mass limit combined with 
parton-hadron duality, formulated as an operator-product expansion. 

On the other side there are the exclusive non--leptonic decays, where 
no theoretically solid basis for a calculation of branching ratios 
exists. However, these decays are of prime interest with respect to 
CP violation and the determination of the CKM matrix. In this 
field the heavy mass limit has not brought any significant progress. 

In this work we have set up a QCD based description for one-particle inclusive 
decays. The basic ingredients are the heavy mass limit and 
a short distance expansion. We obtain operators similar to the ones
describing heavy quarkonia production or one particle inclusive processes.  
We have formulated this method for decays of the type 
$B \to \bar DX$ where $X$ in principle can be any state. 

We have applied this method to one-particle inclusive semi--leptonic decays 
of the form $B \to \bar D X \ell^+ \nu$, studying the leading order 
in the operator-product expansion. Higher order terms are 
suppressed by inverse powers of a large scale related to the 
heavy quark masses. To leading order, all these decays 
are parametrized in terms of eight functions $\eta_i$ and $\rho_i$
which depend on the 
velocities of the $B$ and the $\bar D$ meson. 

The main problem is to obtain these non-perturbative functions 
$\eta_i$ and $\rho_i$ and 
we employed the fact that the inclusive semi--leptonic decays are dominated 
by the two channels $B \to \bar D \ell^+ \nu$ and 
$B \to \bar D^* \ell^+ \nu$. Using this as a starting point we may 
obtain four of the unknown functions (the $\eta_i$) in terms of 
the Isgur-Wise function with a well motivated factorization--ansatz. 
The remaining four (the $\rho_i$) 
are suppressed by powers of $\alpha_s$ as well as 
by factors $1/N_C$. 

Estimating the four functions $\rho_i$ to be of the 
order $1/N_C$ we are able to 
describe the features of the one-particle inclusive semi--leptonic decays. In 
particular, QCD radiative corrections induce a relatively large amount 
of decays which originate not from the exclusive modes 
$B \to \bar D \ell^+ \nu$ and $B \to \bar D^* \ell^+ \nu$. This is 
in accordance with the experimental data giving us some confidence in 
our method. 

The approach suggested in the present paper opens the door to a QCD
based description of one-particle inclusive processes; it is not limited to 
semi--leptonic decays. In particular, the non-perturbative functions
$\eta_i$ and $\rho_i$ are universal and should also describe 
other one-particle inclusive processes.

\section*{Acknowledgements}
We are grateful to Xavier Calmet for producing the plot and for 
performing the numerical analysis. This work was supported by   
the ``Graduiertenkolleg: Elementarteilchenphysik and Beschleunigern'' 
and the ``Forschergruppe: Quantenfeldtheorie, Computeralgebra und 
Monte Carlo Simulationen'' of the Deutsche Forschungsgemeinschaft.

\end{document}